# From Genomes to Algorithms: Neural Network Applications for Palimpsest Detection in Medieval Manuscripts

[1] James B. Harr III, [2] Madelin E. Blong, [3] Tessa Gadomski, [4] Kelly A. Meiklejohn, [5] William E. Gundling Jr.

[1] Assistant Teaching Professor, College of William and Mary, Williamsburg, Virginia USA
[2] Molecular Laboratory Technician, Naveris Inc., Durham, North Carolina USA
[3] Conservation Librarian, University of Pennsylvania Libraries, Philadelphia, Pennsylvania USA
[4] Associate Professor of Forensic Science, Western Sydney University, Sydney, Australia
[5] Associate Professor of Biology, Christian Brothers University, Memphis, Tennessee USA

---

**Abstract:** Biocodicology, the study of biological information preserved in manuscripts, offers new opportunities to examine parchment as both a textual and biological artefact. This study applies non-destructive sampling to isolate and sequence mitochondrial genomes (mtGenomes) from a 14th-century manuscript, Ms. Codex 1629, which contains both single-use and palimpsested folios. We sought to evaluate whether palimpsest preparation, including chemical washing, compromised DNA integrity and whether computational methods could aid in identifying reused parchment. DNA sequencing revealed that both single-use and palimpsested parchments retained sufficient mtGenomes for analysis, with no significant differences in genome coverage or depth. To assess the potential of computational biology in manuscript studies, we implemented machine learning classifiers, including logistic regression and neural networks, to distinguish palimpsests from single-use folios. Models achieved high precision but exhibited reduced recall for the minority palimpsest class, reflecting dataset imbalance. While additional ancient mtGenome samples from palimpsest are required and further testing is needed, this study demonstrates how integrating molecular biology and neural networks highlights new approaches for palimpsest detection and underscores the evolving role of data science in biocodicology.

---

## I. INTRODUCTION

A subset of bioinformatics, biocodicology is an emerging discipline that extracts and analyzes biological information stored and preserved in the organic materials found in parchment (animal skin used as substrate). Integrating codicology (the study of books), molecular biology, and data science, biocodicology embraces interdisciplinarity. Application of biocodicology aids an understanding of processed animal skins by considering them as biological artifacts preserving proteins and genetic materials such as mitochondrial genomes (mtGenomes) and allows for non-invasive, non-destructive sampling to study production, distribution, reuse, and conservation.

Advancements in high-throughput sequencing and bioinformatics have allowed for increasingly robust techniques and approaches to identify and recover from organic cultural heritage materials such as medieval parchment without damaging the artifact. A method developed by Scheible et al. (2024) demonstrated that mitochondrial DNA can be extracted from parchment using a non-invasive brushing technique. Their research demonstrated that a combination of hybridization capture, Illumina sequencing, and a four-tier authentication strategy can be used to identify premodern DNA signals while not damaging the manuscript or compromising parchment integrity. As ancient DNA can be highly fragmented and influenced by noise and contamination, developed strategies to analyze degraded samples are important. Renoud et al. (2015) noted that pipelines such as schmutzi. MapDamage, and PMDtools are needed to correct for damage while estimating levels of contamination, particularly when reconstructing endogenous consensus sequences. As a result, downstream analyses that use PCA, phylogenetic trees, admixture models, etc. allow for sequence clustering and population affinities, allowing research to build the scope of interpretation beyond species identification.

A separate approach created by Fiddyment et al. (2015) uses eZooMS for collagen peptide mass fingerprinting and combines proteomics and genomics as a way to successfully identify parchment species despite DNA degradation. Naderi et al. (2007) and Daly et al (2018) take a broader approach by using phylogeographical surveys to provide haplotypes and diverse lineage baselines as referential data. Their work allows organic samples to be compared by positioning them within the purview of animal domestication and parchment production, linking variations in genetics to cultural and geographical context.

In this study, focus was placed on manuscripts containing palimpsests. A palimpsest is a previously-used substrate that has had all text removed – either by scraping the surface of the substrate or by removing all ink through chemical washing – and then reused as a "blank" surface for new text. This was common practice as preparing animal skin for manuscript use was time-consuming and costly. In short, erasure for repurposing was an economical way to recycle substrates. As questions arose asking whether chemicals used to erase text were destructive, one of our goals was to investigate whether the integrity of the mitochondrial DNA could withstand the chemical washing process while concurrently initiating new conversations on parchment reuse.

Samples for this research were collected from UPenn Ms. Codex 1629, *Commenaria ad Rhetoricam Ciceronis* (University of Pennsylvania Kislak Center for Special Collections), a 14th-century manuscript consisting of both single-use and palimpsested parchments. In addition to the mixture of parchment types, Codex 1629 also includes generous margins and blank space which allowed for testing to take place without intruding on text or ink. Finally, due to the manuscript's provenance (Northern Italy), the likelihood that an inclusion of different species was high.

Computational methods, particularly machine learning (ML), provide powerful tools for addressing classification challenges. By training classifiers on genetic data extracted from parchment, it could be possible to distinguish between single-use and palimpsested folios.

This study applies both conventional statistical analysis and supervised ML models to mtGenomes sequencing data from Codex 1629. Neural network applications were performed in Python using Google Colab, which enabled flexible experimentation with classification, oversampling techniques (SMOTE), and dimensionality reduction (PCA). While models showed strong precision, palimpsests remained harder to detect due to class imbalance (largely due to the limited availability of palimpsest samples), underscoring both the promise and limitations of ML in cultural heritage contexts.

This research demonstrates new tools and approaches for palimpsest identification and characterization, and tests whether computational biology can employ neural networks to limited, noisy, and fragmented genetic materials, further revealing biocodicology as an evolving field with a growing number of methodological approaches.

## II. METHODS

### *Sampling*

Sample locations were chosen after close examination of the manuscript with consideration for its current condition and long-term preservation. The manuscript is fully covered in a limp vellum binding and is sewn on three supports with endbands composed of a plain, undyed cord. While there are paper endsheets and pastedowns, the textblock is otherwise composed of parchment. The text is written in a range of dark brown-black and green-black inks, with significant margins on each page. There are decorative initials in red ink throughout, as well as numerous manicules and annotations in the margins of the text. Overall, the binding is stable and in good condition, but it is not original to the text. The bound volume is slightly wedge-shaped with a gaping fore-edge. There is embedded grime on all exterior surfaces of the binding, as well as minor localized stains. The leaves exhibit minor planar distortions, embedded grime, localized staining, and minimal insect damage. There are minor edge tears and losses throughout, and the text is faded on some pages. Numerous inscriptions in the gutter are partially obscured, and others are close to the fore edge. The red initials show evidence of bleeding and off-setting to adjacent leaves, which suggests previous exposure to high humidity or ambient moisture. There is no visible evidence of previous conservation treatment that might have introduced protein-based sources of contamination.

With these features in mind, samples were primarily taken from the bottom margin of the leaves, avoiding the gutter, which is likely to accumulate dirt and debris, and avoiding the corners and edges of the leaves, which are most likely to be handled by patrons.

99 samples – 52 single-use and 47 palimpsest – were collected from all 26 bifolios, as well as from both the front and back flyleafs. While every sheet of parchment was sampled, priority was given to samples which, after cursory inspections using both raking and multispectral light, were suspected of being palimpsests. All sample collection followed the same standard operating procedure (SOP): sterile latex gloves were frequently replaced and surgical grade face masks were used to avoid peripheral contamination, EndoCervix brushes collected genetic material by brushing the substrate surface for a minimum of 60 seconds in a surface area of 2.5cm x 10cm. (Scheible et al. 2024). Once the samples were collected, the EndoCervix brush heads were detached and transferred to a 2 ml DNA LoBind microcentrifuge tube and sent to the North Carolina State University College of Veterinary Medicine for DNA isolation and sequencing. After collection, the manuscript was reinspected. Overall, the binding, sewing, and parchment remain intact and consistent with the manuscript's condition prior to sampling.

### *DNA isolation*

DNA isolation was performed in four batches of 16 or fewer samples (consisting of 15 parchment samples and one reagent blank) using a previously published method by Teasdale et al. (2017) and Yang et al. (1998). Only 48 out of the 99 total samples were processed, 22 palimpsest and 26 single-use parchments. A total of 750 µl of isolation buffer was added to each sample to fully submerge the brush head and allowed to incubate at 50 ºC with 800 rpm for approximately 24 hours. The concentration of total DNA recovered was determined using the Qubit dsDNA High Sensitivity Assay Kit (ThermoFisher Scientific, Waltham, MA) on the Qubit Fluorometer 3.0 (ThermoFisher Scientific). DNA extracts were stored at -20 ºC until use.

*Library construction*

Isolated DNA were prepared into libraries using the KAPA Hyper Prep Kit (Roche, Basel, Switzerland) as previously described by Scheible et al. (2024) with the following modifications: (A) KAPA unique dual-indexed adapters manufactured by Roche were used to individually barcode (or index) each sample for sequencing, (B) KAPA Pure Beads (Roche) were used for all bead purifications, (C) individual libraries were quantified using the Qubit dsDNA HS Assay Kit (ThermoFisher Scientific) on the Qubit Fluorometer 3.0 (ThermoFisher Scientific), (D) a single pool of all libraries (n = 67) was prepared with a maximum of 20 µl of any library added to the pool, and (E) pooled libraries were purified using a 1.8 X bead ratio of KAPA Pure Beads (Roche) and concentrated to 25 µl.

*Mitochondrial genome (mtGenome) enrichment and sequencing*

mtGenome enrichment and sequencing were performed following the protocol from Scheible et al. (2024) with the following modifications: (A) hybridization capture was performed following the myBaits Hybridization Capture for Targeted NGS- Manual v. 5.02- High Sensitivity Protocol with commercially available cow, goat and sheep mitochondrial genome baits to enrich the source animal mtGenome (Daicel Arbor Biosciences, Ann Arbor MI), (B) two rounds of enrichment were performed at 60 ºC for ~48 hours according to the myBaits protocol, and (C) the concentration of the enriched, amplified pooled library was measured with the Qubit dsDNA HS Assay Kit (ThermoFisher Scientific). The final, enriched library was sequenced using Illumina chemistry on the Miniseq System (Illumina, San Diego, CA) following the protocol from Scheible et al. (2024).

*Data analysis*

Raw sequence reads were imported into the CLC Genomics Workbench (Qiagen) and processed through the workflow established by Scheible et al. (2024) utilizing both merged and unmerged reads in the workflow.

## III. RESULTS

The goal of this project was to determine if sufficient mitochondrial DNA from the source animal could be recovered and sequenced from palimpsested parchments compared to single-use parchments using a nondestructive sampling method. In this study we processed a total of 22 palimpsested and 26 single-use parchment sheets from a single book, Codex 1629. We used two of the metrics outlined in Scheible et al. (2024) for assessing the quality of the resulting mtGenome sequences: (1) percentage of the mtGenome covered, and (2) mean read depth across the mtGenome.

**Table 1. Average mtGenome sequencing metrics from palimpsest (n = 22) and single-use (n = 26) samples collected from Codex 1629. bp, base pairs; SD, standard deviation; #, number.**

| | Consensus length (bp) | Mapped reads (#) | mtGenome coverage (%) | Mean read depth (± SD) |
|---|---|---|---|---|
| Palimpsest (n = 22) | 16,504 | 86,304 | 99 | 574 ± 235 |
| Single-Use (n = 26) | 15,925 | 75,141 | 96 | 521 ± 172 |

There were no significant differences between the palimpsested parchments and the single-use parchments in either the coverage of the mtGenome (size is 16,616 bp) or mean read depth ($p = 0.0799$ and $p = 0.8397$, respectively) (**Table 1**). The threshold specified by Scheible et al. (2024) for genome coverage was 90%. When applying this threshold, all 22 palimpsested samples were above 90% and 23 out of 26 single-use samples were above 90%. The threshold specified by Scheible et al. (2024) for mean read depth is >10 X. All 22 palimpsested samples were above this threshold and 23 out of 26 single-use samples were below. The same three single-use samples did not meet either threshold.

**(A)**

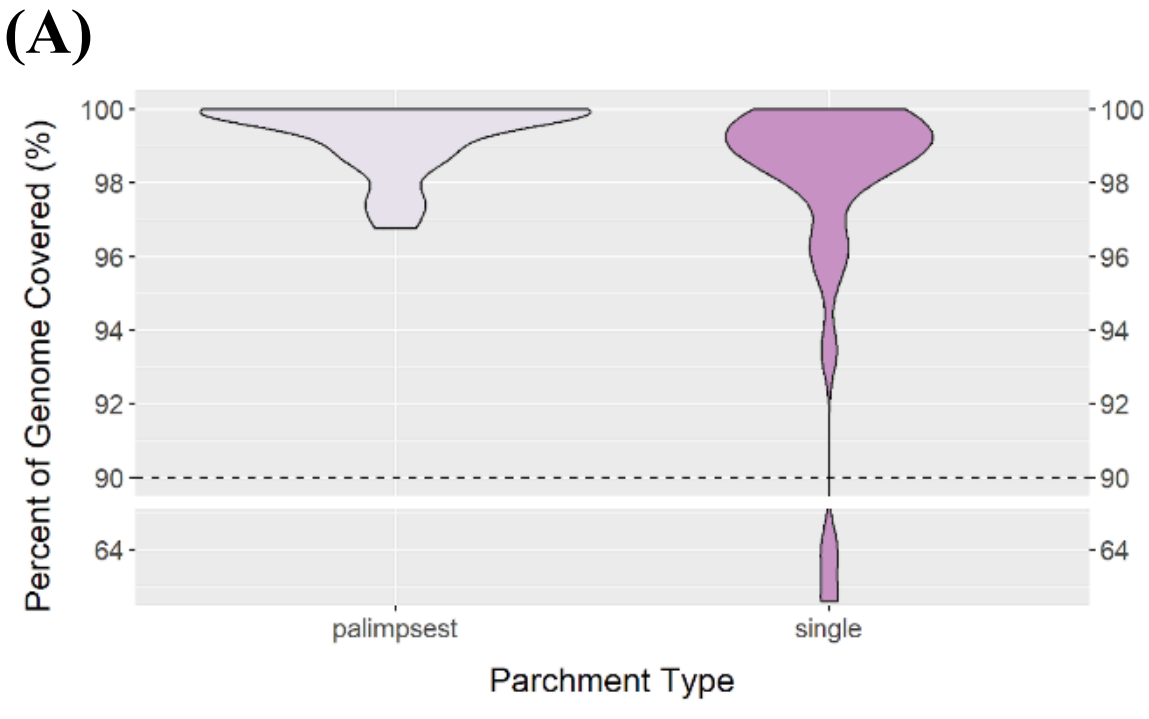

**(B)**

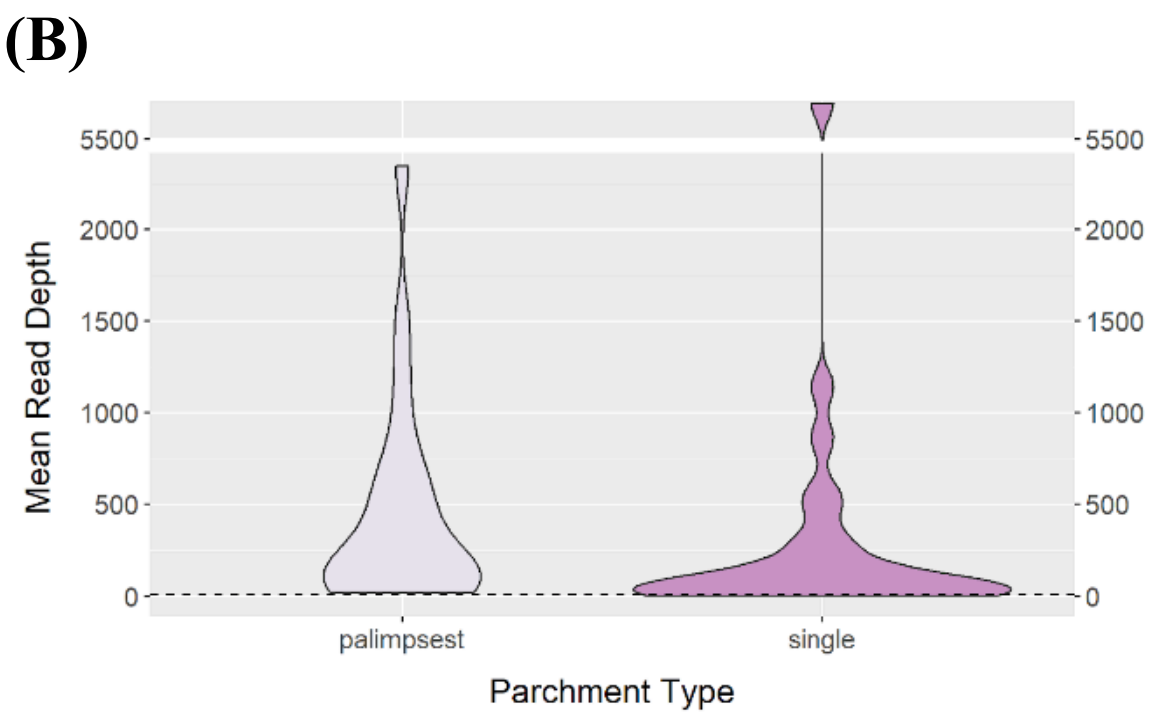

**Figure 1. Violin Plots Comparing Parchment Type to Sequencing Metrics. Violin plots were generated to compare the parchment type and the percent of mtGenome covered (A) and the mean read depth (B). A violin plot is used to visualize data distribution with width indicating data density. The horizontal dashed lines on the plots indicate the thresholds for each metric as identified by Scheible et al. (2024). The light purple represents the palimpsested parchments and the dark purple represents the single-use parchments.**

The consensus mtGenome for each of the samples was searched against GenBank via a default *blastn*. All samples, except one single-use sample had a best match to *Ovis aries* (sheep) with 100% query coverage, E-value of 0, and >99% identity. The one single-use sample mapped to *Capra hircus* (goat).

The results show that sufficient genetic material for sequence-based analysis of the mtGenome is retained on palimpsested parchments (even after chemical washing) that can be isolated using a nondestructive sampling technique. Overall, we identified that the palimpsested parchments and the single-use parchments had similar sequencing metrics.

## IV. MACHINE LEARNING (ML) APPLICATIONS

*Balancing and Testing*

Following the generation of mtGenome consensus sequences, a Python-based neural network was built to test whether an algorithm could identify single-use and palimpsested samples. Approximately 500 external ancient mtGenome *Capri hircus* (goat) and *Ovis aries* (sheep) samples were pulled from GenBank (Cassidy et al. 2017; Meadows et al. 2007; Naderi et al. 2007), randomized, and tested against the Codex 1629 samples. Additionally, to reduce excessive training/testing noise and computational time, the GenBank mtGenome samples were trimmed to 16042-16600 bp to isolate the displacement loop (d-loop) in order to increase comparison consistency. As mtGenome samples from palimpsests are limited to the results of this research, the majority were samples that were either extracted both from parchment that was identified as single-use (single-use) or from parchment that was neither identified as palimpsests nor tested for palimpsesting (untested), creating an inherent class imbalance. Randomized testing and training sets were limited to 124 samples (palimpsests: n=5; single-use/untested: n=119). As a result, the confusion matrix **(Figure 2)** for the test set showed perfect classification of single-use samples (119/119 correctly identified) but partial misclassification of palimpsests; 3 were correctly identified and 2 incorrectly labeled as single-use, reflecting the model's high overall accuracy but reduced sensitivity to the minority palimpsest class **(Table 2)**.

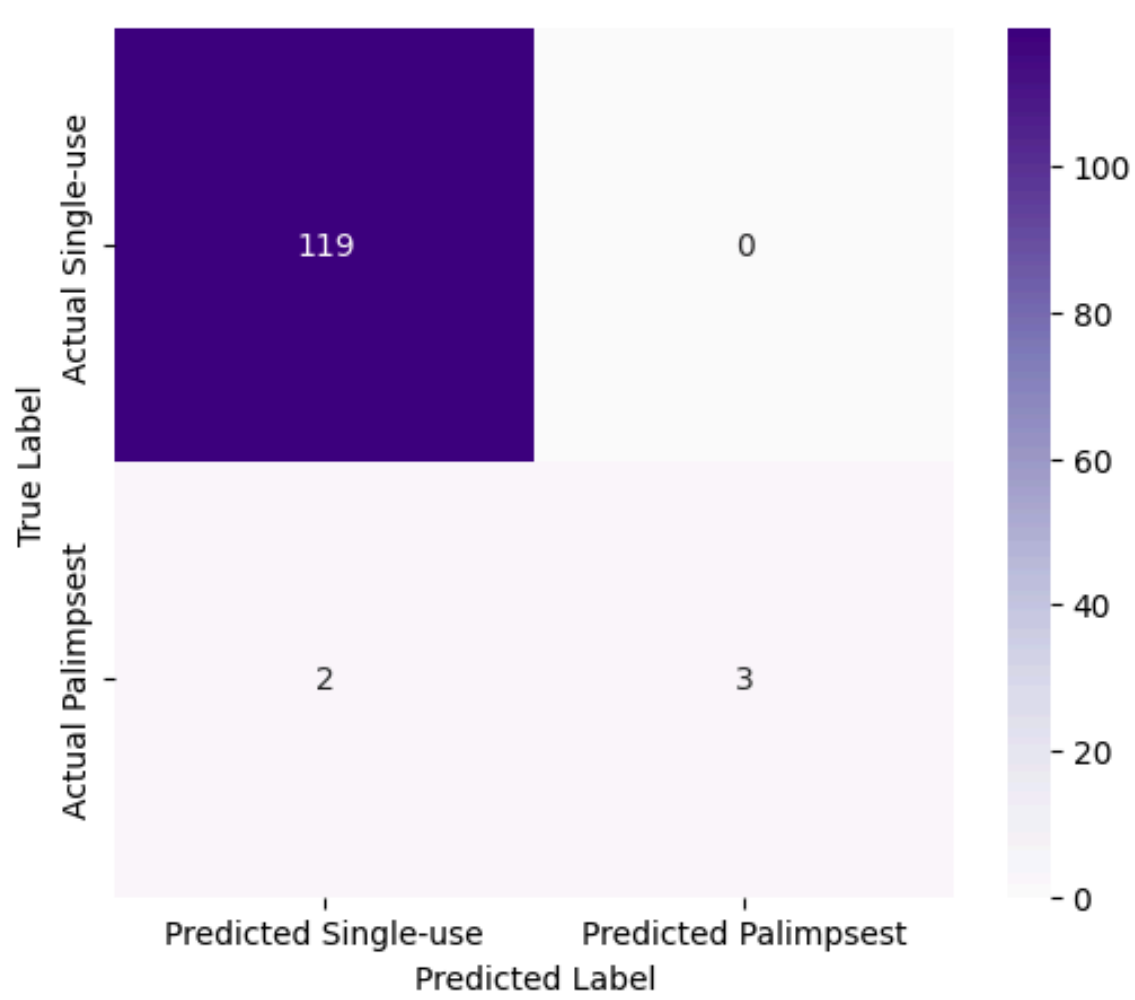

**Figure 2. Balanced classification confusion matrix.**

**Table 2. Classification Report (Test Set)**

| | **Precision** | **Recall** | **F1-Score** |
|---|---|---|---|
| Palimpsest (n=5) | 1.00 | 0.60 | 0.75 |
| Single-use (n=119) | 0.98 | 1.00 | 0.99 |
| Accuracy (n=124) | | | 0.98 |
| macro avg (n=124) | 0.99 | 0.80 | 0.87 |
| weighted avg (n=124) | 0.98 | 0.98 | 0.98 |

Results from principal component analysis (PCA) show clustering where the largest variations occur. PCA visualization of the balanced training set **(Figure 3)** highlights the aforementioned difficulty of separating palimpsest from single-use samples based on k-mer features.

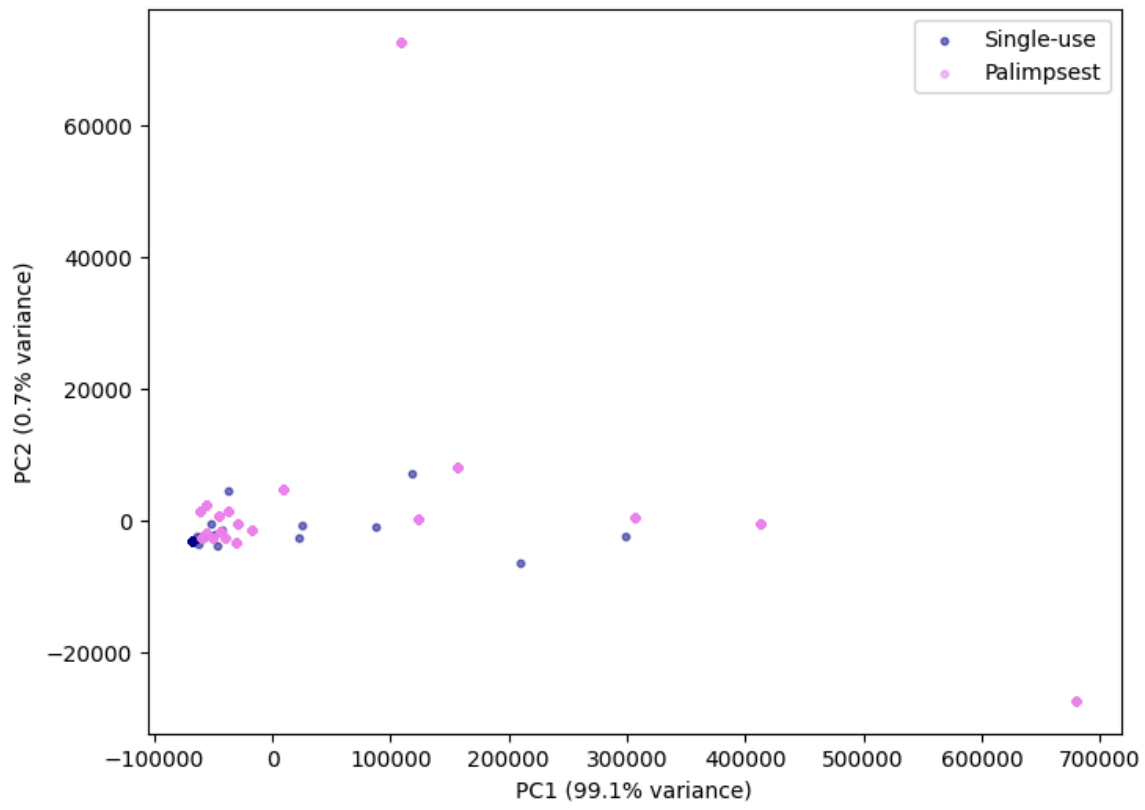

**Figure 3. PCA visualization of balanced training results.**

While PC1 explains the overwhelming majority of variance (99.1%), the overlap between the two classes is notable, indicating that the features extracted from mtGenome sequences do not provide strong discernment for palimpsest detection. The overlap also aligns with the classifier's performance – high overall accuracy (98%), perfect precision for palimpsests, but reduced recall (0.60) – making the model highly conservative in labeling samples as palimpsests. The overlap also avoids false positives yet fails to detect some true palimpsests. Consequently, the PCA distribution underscores the central challenge of this study: imbalanced datasets combined with weak feature separability. Most notably, PCA of sample balancing confirmed classifier bias towards a majority of class optimization. In short, the distribution limits the sensitivity of machine learning models even after analyzing >3 million samples with synthetic minority oversampling technique (SMOTE) **(Figure 4)**.

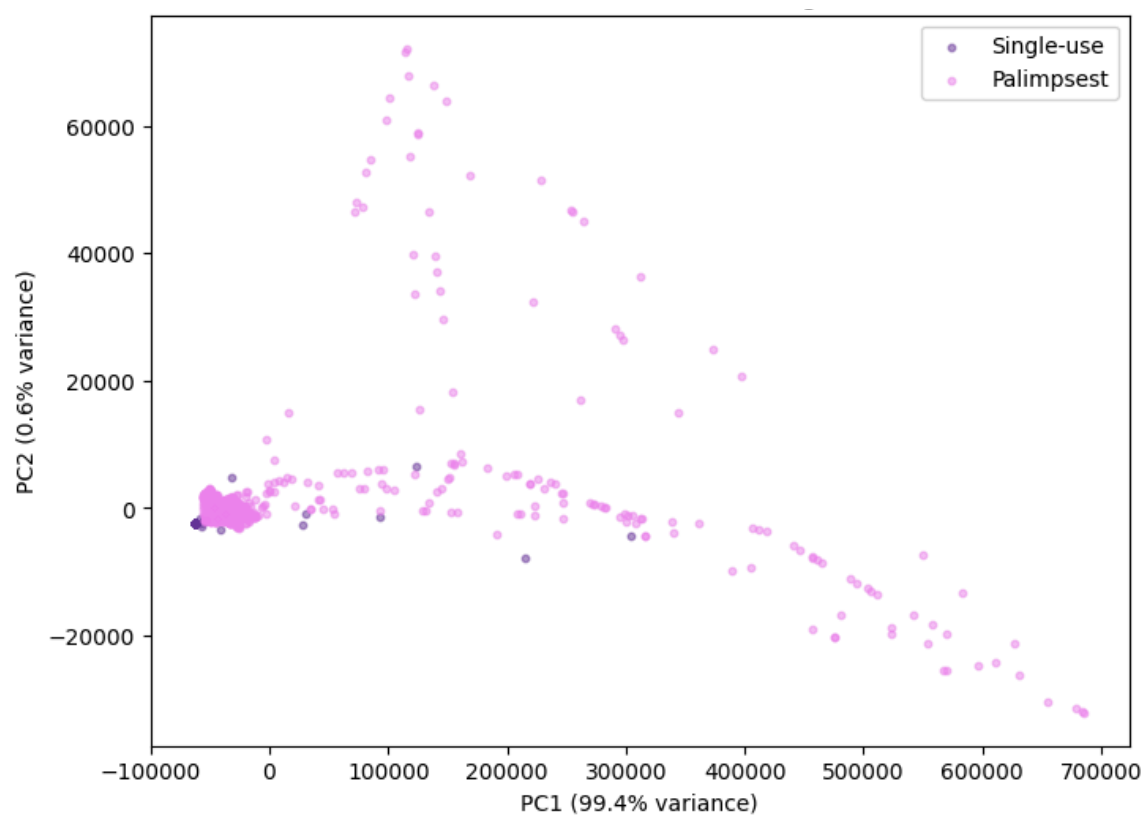

**Figure 4. PCA of SMOTE-balanced training data.**

The PCA of the SMOTE-balanced training data revealed oversampling and altered distribution of palimpsest samples in feature space. After synthetic data generation, palimpsests are far more numerous and spread widely along PC1, which captures 99.4% of the variance. This synthetic expansion improves class balance but also introduces a dispersed and less biologically cohesive cluster of palimpsest points compared to the single-use samples, which remain tightly grouped near the origin. While SMOTE reduces the class imbalance and enhances the model's ability to learn palimpsest-like patterns, the extensive overlap and diffuse distribution highlight the challenge of feature separability. The visualization reinforces why classification recall for palimpsests improves only modestly: oversampling provides more training data, but the inherent similarity of genetic features between single-use and palimpsest parchments still limits distinct separation.

*Sample Augmentation*

As a way to further balance disproportionate single-use/palimpsest sampling, augmentation was used to create random sub-sequences from each of the 24 palimpsest samples rather than full duplicates which would blatantly influence and distort testing results, while allowing data leakage. By applying sequence-level cross-validation and keeping segments from the same sequence in the same fold, data leakage could be avoided. Furthermore, random sub-sequence testing also takes minor biologically plausible variations such as mutations and reverse complements into consideration and simultaneously increasing diversity.

To verify the success of this approach, 538 single-use and 538 palimpsest sample segments were created and trained/tested. The results were excellent (balanced accuracy=0.989) **(Table 3)**. A caveat, however, is that without more independent palimpsest samples, these results, while encouraging, remain largely inconclusive.

**Table 3: Reclassification results following augmentation (538 single-use vs. 538 palimpsest sample segments). Results were excellent with a balanced accuracy of 0.989.**

| | **Precision** | **Recall** | **F1-Score** |
|---|---|---|---|
| Palimpsest (n=538) | 0.978 | 1.000 | 0.989 |
| Single-use (n=538) | 1.000 | 0.978 | 0.989 |
| Accuracy (n=1076) | Very high | Very high | Excellent |

## V. CONCLUSION

This study demonstrates that mtGenomes can be reliably recovered from both single-use and palimpsested parchments using non-destructive sampling, even when chemical washing was historically applied. Sequencing results revealed no significant differences in genome coverage or read depth between parchment types, supporting the biological resilience of palimpsests. Building on these findings, we explored machine learning as a means of manuscript classification, applying logistic regression and neural networks to distinguish reused parchment. Although high precision was achieved, recall for palimpsests remained limited, reflecting dataset imbalance and the rarity of these artifacts.

By conducting all neural network applications in Python on Google Colab, we highlight the accessibility of machine learning for interdisciplinary research, lowering barriers for collaboration between manuscript scholars, molecular biologists, and data scientists. Future research should focus on expanding palimpsest datasets, exploring deep learning architectures, and refining feature engineering to improve model sensitivity. Ultimately, this project underscores the potential of combining molecular biology with computational methods to enhance our ability to detect, analyze, and preserve the complex histories of medieval manuscripts.

While the results are encouraging, further research and expanded success for palimpsest identification using the ML model requires additional ancient/premodern mtGenome samples to balance modern sequences and a more robust catalogue of palimpsest samples.

## ACKNOWLEDGMENTS

We sincerely thank Melissa Scheible (Research Associate, College of Veterinary Medicine, North Carolina State University) for processing and analyzing the genetic samples at the NCSU College of Veterinary Medicine, and reviewing both the data and the article, Dot Porter (University of Pennsylvania Libraries) and Sarah Reidell (University of Pennsylvania Libraries) for all their help accessing and sampling *UPenn Ms. Codex 1629*, Hussin Ragb (Department of Electrical and Computer Engineering, Christian Brothers University) for feedback and assistance with neural network functionality, and the Rosa Deal School of Arts, Christian Brothers University for additional travel resources.

This work was funded by the North Carolina State University, College of Veterinary Medicine and North Carolina State University 2024-2025 Interdisciplinary Undergraduate Research and Mentorship Award.

★★★